\documentclass[conference]{IEEEtran}
\usepackage{amsmath, amsthm, amsfonts, amssymb, amscd}
\usepackage{enumerate}
\usepackage{graphicx}
\usepackage{algorithm}
\usepackage{algorithmic}
\usepackage{booktabs}
\usepackage{siunitx}
\usepackage{seqsplit}
\usepackage{cite}

\ifCLASSOPTIONcompsoc
 \usepackage[caption=false,font=normalsize,labelfont=sf,textfont=sf]{subfig}
\else
 \usepackage[caption=false,font=footnotesize]{subfig}
\fi

\begin{document}
\title{LASANA: \underline{La}rge-Scale \underline{S}urrogate Modeling for \underline{A}nalog \underline{N}euromorphic \underline{A}rchitecture Exploration\vspace{-8pt}}

\author{\IEEEauthorblockN{Jason~Ho,~James~A.~Boyle,~Linshen~Liu,~and~Andreas~Gerstlauer}
\IEEEauthorblockA{\textit{Electrical~and~Computer~Engineering,~The~University~of~Texas~at~Austin},~Texas,~USA}
\{jason\_ho, james.boyle, linshen\_liu, gerstl\}@utexas.edu
}

\maketitle

\begin{abstract}
Neuromorphic systems using in-memory or event-driven computing are motivated by the need for more energy-efficient processing of artificial intelligence workloads. Emerging neuromorphic architectures aim to combine traditional digital designs with the computational efficiency of analog computing and novel device technologies. A crucial problem in the rapid exploration and co-design of such architectures is the lack of tools for fast and accurate modeling and simulation. Typical mixed-signal design tools integrate a digital simulator with an analog solver like SPICE, which is prohibitively slow for large systems. By contrast, behavioral modeling of analog components is faster, but existing approaches are fixed to specific architectures with limited energy and performance modeling. In this paper, we propose LASANA, a novel approach that leverages machine learning to derive data-driven surrogate models of analog sub-blocks in a digital backend architecture. LASANA uses SPICE-level simulations of a circuit to train ML models that predict circuit energy, performance, and behavior at analog/digital interfaces. Such models can provide energy and performance annotation on top of existing behavioral models or function as replacements to analog simulation. We apply LASANA to an analog crossbar array and a spiking neuron circuit. Running MNIST and spiking MNIST, LASANA surrogates demonstrate up to three orders of magnitude speedup over SPICE, with energy, latency, and behavioral error less than 7\%, 8\%, and 2\%, respectively.
\end{abstract}


\section{Introduction}
\IEEEPARstart{W}{ith} the exponential growth in machine learning (ML) complexity and model size, the memory bottleneck and energy consumption of traditional von Neumann computation have become key concerns. Neuromorphic computing is a promising alternative that draws on brain-inspired principles to enable in-memory and/or event-driven computation. A subset of neuromorphic systems seeks to exploit analog computation for its improved energy-efficiency~\cite {joubert_hardware_2012} and ability to leverage the unique properties of novel devices~\cite{kim_emerging_2020}. For instance, ML architectures are replacing matrix-vector multiplication with efficient compute-in-memory (CiM) analog multiply-accumulate operations using memristive crossbars~\cite{amin_imac-sim_2023, shafiee_isaac_2016, cosemans_towards_2019}. Analog spiking systems are another emerging paradigm offering event-driven processing, enabling parallel computation and further energy savings \cite{pehle_brainscales-2_2022, ankit_resparc_2017}.

\begin{figure}[t!]
    \centering
    \includegraphics[width=0.485\textwidth]{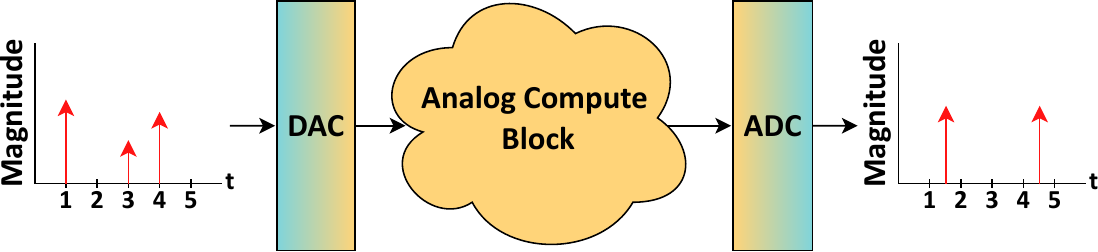}
    \caption{Hybrid architecture with a digital backend and analog compute cores.}
    \label{fig:hybrid}
    \vspace{-5pt}
\end{figure}

In these architectures, analog compute blocks are typically integrated into a digital backend that handles control logic and supports scalability. Each block interfaces with the digital world through converters that translate between the continuous analog and discrete digital domains, as shown in Fig.~\ref{fig:hybrid}. Since the analog block lies on the critical path, its energy consumption and delay are important to the overall energy efficiency and performance of a design. At the same time, overall performance and efficiency are determined by the integration of analog and digital domains, necessitating their joint exploration.

A crucial problem in the co-design of such systems is the lack of tools that allow for fast and accurate modeling and simulation of large-scale hybrid analog/digital architectures. Traditional approaches for modeling analog components rely on low-level simulation methods such as SPICE, which is prohibitively slow for large systems. Early-stage SoC-scale simulations are often accomplished with hand-crafted behavioral models like SystemVerilog Real Number modeling (SV-RNM) and SystemC-AMS~\cite{scherr_beyond_2020}, which replace complex transient calculations with simplified mathematical models evaluated at discrete events. However, manual annotation of energy and performance estimates is required. Some custom-developed behavioral simulators, e.g.\ for analog crossbar arrays~\cite{lee_system-level_2019,plagge_athena_2022,CrossSim} include analytical energy and performance models, but they are limited in accuracy and specific to one type of architecture, restricting their flexibility.

In this paper, we propose LASANA, a novel data-driven approach using machine learning to derive lightweight analog surrogate models that estimate the energy, latency, and behavior of analog sub-blocks at event-based analog/digital interfaces. Given a circuit netlist, we introduce an end-to-end framework that generates a representative dataset, trains ML predictors, and outputs C++ inference models. Models estimate energy and performance at the granularity of digital clock steps with high accuracy and low overhead. ML-based models can be used to augment existing behavioral models with automatically derived energy and performance estimates or to serve as standalone drop-in surrogates. To the best of our knowledge, this is the first approach to develop ML-based surrogate models of analog circuit blocks at a coarse-grain event level. The key contributions of this work are as follows:

\begin{itemize}
    \item We propose a novel learning formulation and automated approach for the generation and deployment of event-driven ML-based analog surrogate models integrated into digital backend simulators. Our formulation incorporates several optimizations to improve runtime by batching inferences across the system and merging inactive input periods into single events.
    \item We examine runtime and accuracy trade-offs of ML models for coarse-grain analog surrogate modeling. We find that boosted trees and multi-layer perceptrons perform the best depending on circuit complexity. 
    \item We evaluate LASANA modeling on a crossbar array and a spiking neuron circuit. On MNIST and spiking MNIST, we demonstrate up to three orders of magnitude speedup over SPICE, with energy, latency, and behavioral error less than 7\%, 8\%, and 2\%, respectively.
\end{itemize}

\section{Related Work}
Traditionally, analog circuits are simulated with SPICE, which uses numerical methods to solve Kirchhoff's laws. To model mixed-signal designs, tools like Verilog-AMS run SPICE simulations alongside digital simulators~\cite{noauthor_verilog-ams_2014}. SV-RNM and SystemC-AMS are state-of-the-art large-scale behavioral analog simulation methodologies~\cite{scherr_beyond_2020}. They use the digital event solver and simple mathematical equations to model analog circuits, providing simulation speedup at the cost of accuracy. Developing such models requires domain knowledge of the underlying equations of a circuit. Furthermore, they do not provide energy or performance estimation, requiring manual annotation. 

Existing architectural simulators that incorporate analog circuits typically rely on custom-developed simulators~\cite{lee_system-level_2019, plagge_athena_2022, CrossSim}. They combine behavioral modeling with analytical performance and energy estimation of one class of circuit, such as memristor or floating gate crossbar arrays, limiting the range of architectures that can be explored. 

Several works are aimed at replacing SPICE-level transient simulations using ML models targeting one aspect, such as behavioral~\cite{amaral_ann-based_2023} or power~\cite{grabmann_power_2019} estimation. However, they rely on large fine-tuned ML models to provide estimates at fine temporal granularity. By contrast, our approach models behavior, energy, and performance at a coarse-grain event level using fast and simple ML models to support large-scale architecture-level simulation.

\section{Background}
\begin{figure}[t!]
    \centering
    \includegraphics[width=0.485\textwidth]{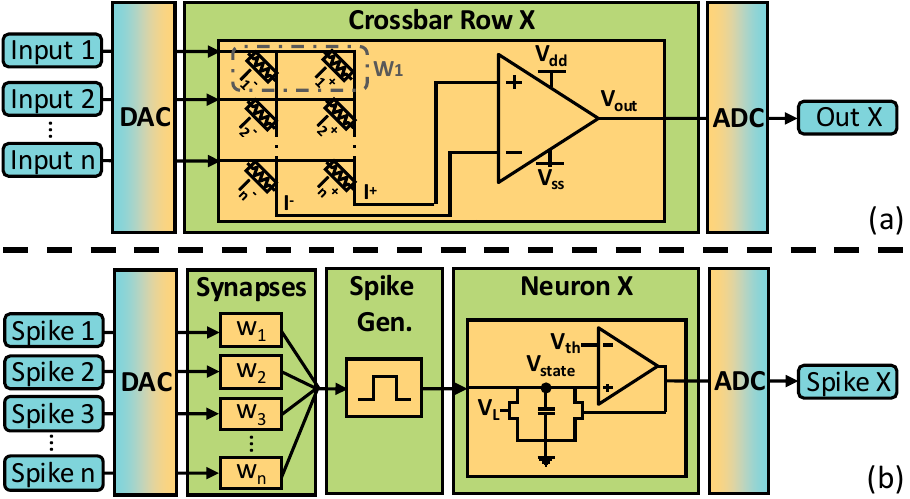}
    \caption{(a) An $n$-length crossbar row, (b) A LIF analog spiking neuron.}
    \vspace{-5pt}
    \label{fig:building-blocks}
\end{figure}

Fig.~\ref{fig:building-blocks} shows two examples of common analog compute blocks in emerging neuromorphic architectures: memristive crossbar arrays and spiking neurons. Fig.~\ref{fig:building-blocks}(a) shows a schematic of one row of a crossbar array embedded in a digital backend. To realize positive and negative weights, each input is connected to two memristors and combined at a differential amplifier, creating an overall weight, $W_i \propto G_{pos,i} - G_{neg,i}$.

Fig.~\ref{fig:building-blocks}(b) shows a schematic of a leaky-integrate-and-fire (LIF) analog neuron that, unlike traditional neural networks, implements spiking neural network (SNN) behavior, which encodes information in the amplitude and timing of discrete spikes sent between neurons. Incoming spikes are modulated by a synapse weight and passed through a spike generator, which converts the voltage into a current spike. Inside the neuron, charge is integrated into a capacitor with an internal state ($V_{state}$). The state is constantly compared against a threshold ($V_{th}$), producing $V_{dd}$ and resetting the circuit state if greater. The state also leaks over time and is implemented with a subthreshold transistor.

Typically, parallel instances of these blocks, each with variations in input and exact circuit parameters, sit between digital-to-analog converters (DACs) and analog-to-digital converters (ADCs) while a digital backend manages the interconnect and control logic of the larger system. Inputs arrive at the analog block at digital timesteps, and the block must finish processing the input before the next clock step.

\begin{figure}[t!]
    \centering
    \includegraphics[width=\linewidth]{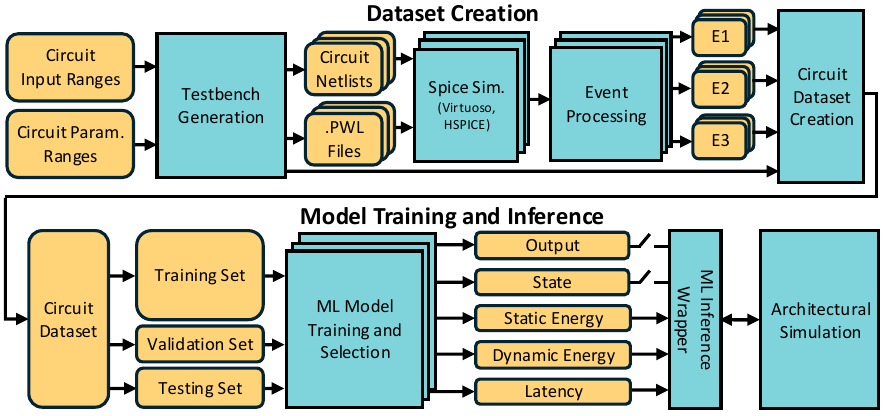}
    \caption{LASANA automated dataset and surrogate model generation flow.}
    \vspace{-5pt}
    \label{fig:our-flow}
\end{figure}

\section{Event-Based Surrogate Modeling}
As shown in Fig.~\ref{fig:our-flow}, our approach develops surrogate models by first creating a representative event-based dataset of a SPICE-level circuit, and then using the dataset to train ML inference models of circuit energy, latency, output, and state behavior. Generated models are wrapped in an inference function that can be plugged into a backend architectural simulator for energy, latency, and optionally behavioral estimation. Energy and latency models can be used to annotate traditional behavioral models, or combined with our ML-based state and output predictors to provide a complete model of analog sub-blocks in a larger digital simulation. In the following, we will describe LASANA modeling steps in more detail.

\subsection{Dataset Creation}
Given a SPICE netlist of an analog circuit, LASANA provides an end-to-end methodology for automated dataset generation to characterize energy, latency, and behavior. To facilitate generalizability, the circuit is treated as a black box, only requiring access to the backend clock, inputs, outputs, state (if applicable), and tunable circuit parameters such as memristive weights or voltage knobs. The dataset is created by running SPICE simulations of a template circuit with randomized inputs and parameters, followed by event processing that splits continuous traces into coarse-grain events.
 
\subsubsection{Testbench Generation} For each SPICE simulation, LASANA randomly generates inputs for a number of simulated timesteps and randomly samples the tunable circuit parameters within their respective ranges uniformly. We define circuit parameters as fixed knobs during simulation (e.g., memristive weights) while inputs vary over time. Each timestep is classified as either active, where inputs change, or static, where inputs remain constant with a probability of $\alpha$ and 1-$\alpha$, respectively. In an active timestep, inputs are randomly sampled within user-defined input ranges. Each run's inputs are written into a .PWL file as a sequence of discrete time-value pairs. At the same time, circuit netlist instances are generated from a template that is modified with the randomized circuit parameters and connected to the inputs.

\subsubsection{SPICE Simulation} Once all inputs have been generated, LASANA runs all netlists with their .PWL files as inputs. All netlists are executed in parallel, using multithreading within each simulation and multiprocessing across the system. 

\subsubsection{Event Processing} 
\begin{figure}[t!]
    \centering
    \includegraphics[width=0.48\textwidth]{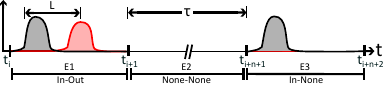}
    \caption{Simulation is discretized to three distinct events (E1, E2, E3). Inputs are in gray, and outputs are in red.} 
    \label{fig:captured-events}
\end{figure}

After simulation, transient data is decomposed into events which always start and end at timestep boundaries, as shown in Fig.~\ref{fig:captured-events}. We classify events into three categories to support event-specific ML formulations and training. E1 and E3 are one-timestep-long events triggered by a change in input and differentiated based on whether there is a change in the output. E2 is a variable-length event that captures any idle period between active timesteps. 

\subsubsection{Circuit Dataset Creation} For each event, LASANA captures the output $o$, the internal state $v_i$ and $v_{i+n}$ of the circuit at the beginning $t_i$ and end $t_{i+n}$ of the event, the length of the event $\tau$, the input values $\mathbf{x}$, and the randomized circuit parameters $\mathbf{p}$ that the simulation ran with. If a circuit is not stateful, the state value is set to 0. Energy $E$ is defined for the duration of the event and classified as dynamic if that timestep produced changes in the output (E1) and static otherwise (E2, E3). Latency $L$ is only calculated for E1 events as the time from the start of the input to the point where the output reaches 90\% of its final value. For spiking signals, latency is defined from the start of the input to the output peak.

\begin{figure}[t!]
    \centering
    \includegraphics[width=0.48\textwidth]{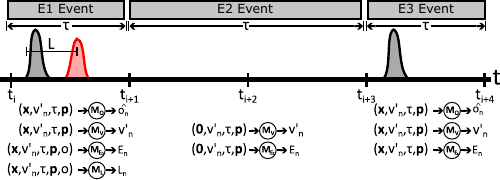}
    \caption{ML transient inference of one analog circuit sub-block.}
    \label{fig:timestep-simulation}
\end{figure}

\subsection{ML Model Training}
As shown in Fig.~\ref{fig:our-flow}, LASANA uses the circuit dataset to train ML models on five separate predictors: an output predictor ($M_O$), a state predictor ($M_V$), a dynamic energy predictor ($M_{E_D}$) that predicts energy per E1 event, a static energy predictor ($M_{E_S}$) that predicts energy for E2 and E3 events, and a latency predictor ($M_L$). We split predictors into separate models to enable training on event-specific data with independent loss optimization. For instance, latency is only meaningful when there is a change in output (E1 events), where including other events would degrade learning.

All predictors take input vector $\mathbf{x}$, current state $v'_i$, the length of the event $\tau$, and circuit parameter vector $\mathbf{p}$ as features. If there is no input, $\mathbf{x}$ is set to 0. Since dynamic energy and latency depend on the output voltage transition, $M_{E_D}$ and $M_L$ also take the previous output $o$ as input. During training, several models are evaluated on a validation dataset, and the best overall ML model for each predictor is selected for inference.

\renewcommand{\algorithmicrequire}{\textbf{Input:}}
\renewcommand{\algorithmicensure}{\textbf{Output:}}
\newcommand{\algorithmicstate}{\textbf{State:}}
\newcommand{\State}{\item[\algorithmicstate]}

\begin{algorithm}[t!]
\caption{ML Inference Wrapper}
\label{alg:function}
\begin{algorithmic}[1]
\footnotesize{
\REQUIRE number of circuits $N$
\REQUIRE current time $t$
\REQUIRE digital clock period $T$
\REQUIRE set $\mathcal{S}$ of circuit IDs with changed input at time $t$
\REQUIRE matrix $\mathbf{X} = (\mathbf{x_0}, \mathbf{x_1}, ..., \mathbf{x_{N-1}})$ of input vectors
\REQUIRE matrix $\mathbf{P} = (\mathbf{p_0}, \mathbf{p_0}, ..., \mathbf{p_{N-1}})$ of circuit parameters
\State vec $\mathbf{t'} = (t'_0, t'_1, ..., t'_{N-1})$ of latest circuit update times
\State vec $\mathbf{v'} = (v'_0, v'_1, ..., v'_{N-1})$ of latest circuit states

\ENSURE vec $\mathbf{e} = (e_0, e_1, ..., e_{N-1})$ of per circuit energy
\ENSURE vec $\mathbf{l} = (l_0, l_1, ..., l_{N-1})$ of per circuit latency
\ENSURE vec $\mathbf{o} = (o_0, o_1, ..., o_{N-1})$ of per circuit output
\STATE $\mathbf{e} = [0, 0, \dotsc, 0], \mathbf{l} = [0, 0, \dotsc, 0]$
\STATE $\mathcal{I} \leftarrow \emptyset$
\STATE \textbf{for} $n \in \mathcal{S}$ \textbf{do} \hfill $\triangleright$ Create idle period batch
    \STATE \hspace{1em} \textbf{if} $t'_n < t-T$ \textbf{then}
        \STATE \hspace{2.4em} $\mathcal{I} \leftarrow \mathcal{I} \cup (0, v'_n, t-t'_n-T, \mathbf{p_n})$
    \STATE \hspace{1em} \textbf{end if} 
\STATE \textbf{end for}

\STATE $\mathbf{\hat{v}'} \leftarrow M_V(\mathcal{I})$ \hfill $\triangleright$ Predict state and static energy
\STATE $\mathbf{\hat{e_S}} \leftarrow M_{E_S}(\mathcal{I})$

\STATE $\mathcal{I} \leftarrow \emptyset$

\STATE \textbf{for} $n \in \mathcal{S}$ \textbf{do} \hfill $\triangleright$ Process outputs and create input batch
    \STATE \hspace{1em} \textbf{if} $t'_n < t-T$ \textbf{then}
        \STATE \hspace{2.4em} $v'_n \leftarrow \hat{v}'_n$
        \STATE \hspace{2.4em} $e_n \leftarrow \hat{e_S}_n$
    \STATE \hspace{1em} \textbf{end if}

    \STATE \hspace{1em} $\mathcal{I} \leftarrow \mathcal{I} \cup (\mathbf{x_n}, v'_n, T, \mathbf{p_n}) $
\STATE \textbf{end for}

\STATE $\mathbf{\hat{e_D}} \leftarrow M_{E_D}(\mathcal{I}, \mathbf{o})$
\STATE $\mathbf{\hat{l}} \leftarrow M_{L}(\mathcal{I}, \mathbf{o})$
\STATE $\mathbf{\hat{o}} \leftarrow M_{O}(\mathcal{I})$   \hfill $\triangleright$ Run all predictors on input batch
\STATE $\mathbf{v'} \leftarrow M_{V}(\mathcal{I})$ 
\STATE $\mathbf{\hat{e_S}} \leftarrow M_{E_S}(\mathcal{I})$

\STATE \textbf{for} $n \in \mathcal{S}$ \textbf{do} \hfill $\triangleright$ Select results based on output behavior
    \STATE \hspace{1em} \textbf{if} $\mathbf{o_n} \neq \mathbf{\hat{o_{n}}}$ \textbf{then}
        \STATE \hspace{2.4em} $e_n \leftarrow e_n + \hat{e_D}_n$
        \STATE \hspace{2.4em} $l_n \leftarrow \hat{l}_n$
    \STATE \hspace{1em} \textbf{else}
        \STATE \hspace{2.4em} $e_n \leftarrow e_n + \hat{e_S}_n$
    \STATE \hspace{1em} \textbf{end if}
    \STATE \hspace{1em} $t'_n \leftarrow t$  \hfill $\triangleright$ Change latest circuit update time
\STATE \textbf{end for}
\RETURN $\mathbf{e}, \mathbf{l}, \mathbf{\hat{o}}$
}
\end{algorithmic}
\end{algorithm}

\subsection{ML Inference}
\label{runtime-sim}

We embed the trained models in a wrapper function for integration with architectural simulators. Which predictors are invoked during inference depends on what events occurred, as shown in Fig.~\ref{fig:timestep-simulation}. $M_V$ is invoked for all events, and $M_O$ is invoked for events with a change in input. $M_L$ and $M_{E_D}$ are only invoked for events with a change in output, while $M_{E_S}$ is invoked otherwise. In the process, we apply several optimizations to improve inference speed, including batching of inferences across the system and merging of inactive input periods into single events.

Algorithm \ref{alg:function} shows the wrapper function for prediction of $N$ analog circuit sub-blocks at time $t$, with a digital clock period of $T$. This function also takes in the set $\mathcal{S}$ of circuit IDs with a change in input at $t$, a matrix $\mathbf{X}$ of the respective inputs to be applied to each circuit, a matrix $\mathbf{P}$ of the $N$ circuit parameters, a vector $\mathbf{t'}$ of the latest state update times for each circuit, and the vector $\mathbf{v'}$ of the current state of each of the $N$ circuits. This function returns the energy $\mathbf{e}$, latency $\mathbf{l}$, and output $\mathbf{o}$ per circuit. If LASANA models are used to provide energy and latency annotation on existing behavioral models, those models would supply the state.

The algorithm first updates the state of all circuits that have a change in input in the current timestep but have not been updated for at least one timestep before. To do so, corresponding input parameters are collected into a batch $\mathcal{I}$ (lines 2-7) that is passed into state and static energy predictors (lines 8-9). Inference results are used to update corresponding state and energy values (lines 13-14) while also collecting the input event batch from the inputs $\mathbf{x_n}$ and parameters $\mathbf{p_n}$ (line 16). The five predictors are called on this batch (lines 18-22), where energy and latency values are computed depending on the output prediction (lines 24-29). Lastly, the latest circuit update time is set to the current timestep $t$ (line 30).

\section{Experiments and Results}
\label{section-5}
We applied our approach to a $32\times32$ memristive crossbar array and an analog spiking neural network implementation. We use LASANA to derive surrogate models for independent 32-input crossbar rows from~\cite{amin_imac-sim_2023} using 1T-1R phase change memory (PCM) bitcells with the PTM HP 14nm library~\cite{sinha_exploring_2012}, and for a 20-transistor spiking LIF neuron from~\cite{indiveri_low-power_2003} using the FreePDK 45nm LP library~\cite{stine_freepdk_2007}, both with a load capacitance of 500\,fF. Circuit parameters included as ML features for prediction include the 32 weights and 1 bias in the crossbar row, and the 4 tunable voltage knobs ($V_{leak}$, $V_{th}, V_{adap}, V_{refrac}$) in the LIF neuron. The crossbar row has a larger input space, but the LIF neuron has stateful, more complex behavior. Behavioral models for both designs were created in SV-RNM and evaluated in Cadence Xcelium 23.03. All runs were performed on a Linux system with a 16-core Intel i7-13700 and 32GB of DDR5 memory. 

\begin{table}[t!]
\centering
\caption{Total model training and testing times.}
\vspace{-5pt}
\label{table:model_times}
\setlength{\tabcolsep}{9pt} 
\begin{tabular}{@{}ccccc@{}}
\toprule
\multicolumn{1}{l}{} & \multicolumn{2}{c}{\textbf{PCM Crossbar Row   \cite{amin_imac-sim_2023}}} & \multicolumn{2}{c}{\textbf{LIF Neuron   \cite{indiveri_low-power_2003}}} \\ \cmidrule(lr){2-3} \cmidrule(l){4-5}
\textbf{Model}       & \textbf{Train (s)}                 & \textbf{Test (s)}                 & \textbf{Train (s)}                  & \textbf{Test (s)}                  \\ \midrule
Mean                 & 0.001                              & 0.0001                            & 0.002                               & 0.0001                             \\
Table                & 0.254                              & 335.3401                          & 0.065                               & 0.1284                             \\
Linear               & 0.340                              & 0.1527                            & 0.073                               & 0.0341                             \\
CatBoost (d=10)             & 26.060                             & 0.0249                            & 11.289                              & 0.0106                             \\
MLP (100,50)                 & 107.049                            & 0.0791                            & 25.931                              & 0.0258                             \\
\bottomrule
\end{tabular}
\end{table}

\begin{table*}[t!]
\centering
\caption{Comparison of ML Model accuracy on the 32-input crossbar row and LIF spiking neuron.}
\vspace{-5pt}
\label{tab:ml-model-analysis}
\setlength{\tabcolsep}{6.8pt}
\begin{tabular}{@{}cccccccccccccc@{}}
\toprule
\multicolumn{1}{l}{} & \multicolumn{6}{c}{\textbf{PCM Crossbar Row \cite{amin_imac-sim_2023}}}                                                                                                    & \multicolumn{7}{c}{\textbf{LIF   Neuron \cite{indiveri_low-power_2003}}}                                                                                                                                   \\ \cmidrule(lr){2-7} \cmidrule(l){8-14}
\multicolumn{1}{l}{} & \multicolumn{2}{c}{\textbf{$M_L$}}        & \multicolumn{2}{c}{\textbf{$M_{E_D}$}}           & \textbf{$M_{E_S}$}               & \textbf{$M_O$}                  & \multicolumn{2}{c}{\textbf{$M_L$}}               & \multicolumn{2}{c}{\textbf{$M_{E_D}$}}           & \textbf{$M_{E_S}$}               & \textbf{$M_V$}                  & \textbf{$M_O$}                  \\ \cmidrule(lr){2-3} \cmidrule(lr){4-5} \cmidrule(lr){6-6} \cmidrule(lr){7-7} \cmidrule(lr){8-9} \cmidrule(lr){10-11} \cmidrule(lr){12-12} \cmidrule(lr){13-13} \cmidrule(l){14-14}        
\textbf{}            & \textbf{MSE}                     & \textbf{MAPE}  & \textbf{MSE}                     & \textbf{MAPE} & \textbf{MSE}                     & \textbf{MSE}                    & \textbf{MSE}                     & \textbf{MAPE} & \textbf{MSE}                     & \textbf{MAPE} & \textbf{MSE}                     & \textbf{MSE}                    & \textbf{MSE}                    \\
\textbf{Model}       & \textbf{(ps\textsuperscript{2})} & \textbf{(\%)}  & \textbf{(fJ\textsuperscript{2})} & \textbf{(\%)} & \textbf{(fJ\textsuperscript{2})} & \textbf{(V\textsuperscript{2})} & \textbf{(ns\textsuperscript{2})} & \textbf{(\%)} & \textbf{(pJ\textsuperscript{2})} & \textbf{(\%)} & \textbf{(pJ\textsuperscript{2})} & \textbf{(V\textsuperscript{2})} & \textbf{(V\textsuperscript{2})} \\ \midrule
Mean        & 68.4                             & 1.558          & 106.2                            & 16.65         & 999.7                            & 0.2184                          & 0.1629                           & 24.50         & 0.131                            & 27.41         & 0.0071                           & 0.0708                          & 0.7070                          \\
Table                & 135.9                            & 2.011          & 190.7                            & 21.44         & 2173.4                           & 0.2742                          & 0.0305                           & 7.18          & 0.011                            & 10.42         & 0.0034                           & 0.0089                          & 0.0292                          \\
Linear               & 68.3                             & 1.557          & 89.3                             & 15.88         & 162.8                            & 0.1952                          & 0.0724                           & 13.86         & 0.041                            & 16.53         & 0.0055                           & 0.0540                          & 0.0283                          \\
CatBoost (d=10)      & \textbf{66.2}                    & \textbf{1.497} & \textbf{2.0}                     & \textbf{1.96} & 35.8                             & \textbf{0.0121}                 & 0.0148                           & \textbf{4.69} & 0.007                            & \textbf{6.33} & \textbf{0.0009}                  & 0.0029                          & 0.0163                          \\
MLP (100,50)         & 67.3                             & 1.563          & 14.6                             & 5.75          & \textbf{34.8}                    & 0.0292                          & \textbf{0.0146}                  & 5.04          & \textbf{0.005}                   & 6.79          & 0.0010                           & \textbf{0.0028}                 & \textbf{0.0161}                 \\ \bottomrule
\end{tabular}
\end{table*}

The crossbar dataset is composed of 1000 random runs of length 500\,ns, simulated in Synopsys HSPICE at 250\,MHz with $\alpha$ = 0.8. Each of the weights and bias were chosen from $w\in\{-1, 0, 1\}$ and 32 inputs from $x \in [-0.8V, 0.8V]$. Dataset generation took 199\,s and resulted in 99,353 E1 and 19,826 E2 events. The LIF neuron dataset is composed of 2000 random runs of length 500\,ns, simulated in Cadence Spectre at 200\,MHz and $\alpha$ = 0.8. Each of the weights and voltage parameters were chosen from $w \in [-1,1]$ and $p \in [0.5V, 0.8V]$, and inputs ranged from $x \in [0V, 1.5V]$ with the number of spikes in each timestep from $n \in [0,5]$. Dataset generation took 403\,s and resulted in 31,263 E1 events, 46,315 E2 events, and 68,009 E3 events. All simulations were parallelized across 20 processes, and datasets were split run-wise into 70\% for training, 15\% for testing, and 15\% for validation.

Table \ref{table:model_times} shows the models evaluated with their total training times and testing times. Models are listed in order of complexity and training time. We compare a boosted tree model and a multi-layer perceptron (MLP) against a linear regressor as well as mean and table-based models that mirror analytical energy and performance estimation in existing behavioral simulators, e.g.\ for crossbar arrays~\cite{plagge_athena_2022}. \textit{Mean} is a constant estimator using the training mean. \textit{Table} is a nearest neighbors estimator similar to table-based models in traditional circuit simulators. Table-based inference time is dominated by expensive distance operations. As input dimensionality increases with the crossbar, inference time suffers. \textit{Linear} minimizes least squares. \textit{CatBoost} is a gradient-boosting model that builds an ensemble of weak decision trees with a max depth of 10, where each new tree iteratively corrects errors of the prior ensemble. The \textit{MLP} is a neural network trained with the Adam optimizer, employing ReLU activation, and two hidden layers of 100 and 50 neurons. Model hyperparameters (learning rate, L2 regularization) were optimized using grid search and trained until the change in validation loss fell below $10^{-5}$. The models were implemented with Scikit-Learn \cite{pedregosa_scikit-learn_2011} and CatBoost \cite{prokhorenkova_catboost_2018}.

\begin{figure}[t]
\vspace{-8mm}
\centering
\subfloat[$M_{E_D}$]{\includegraphics[width=0.248\columnwidth]{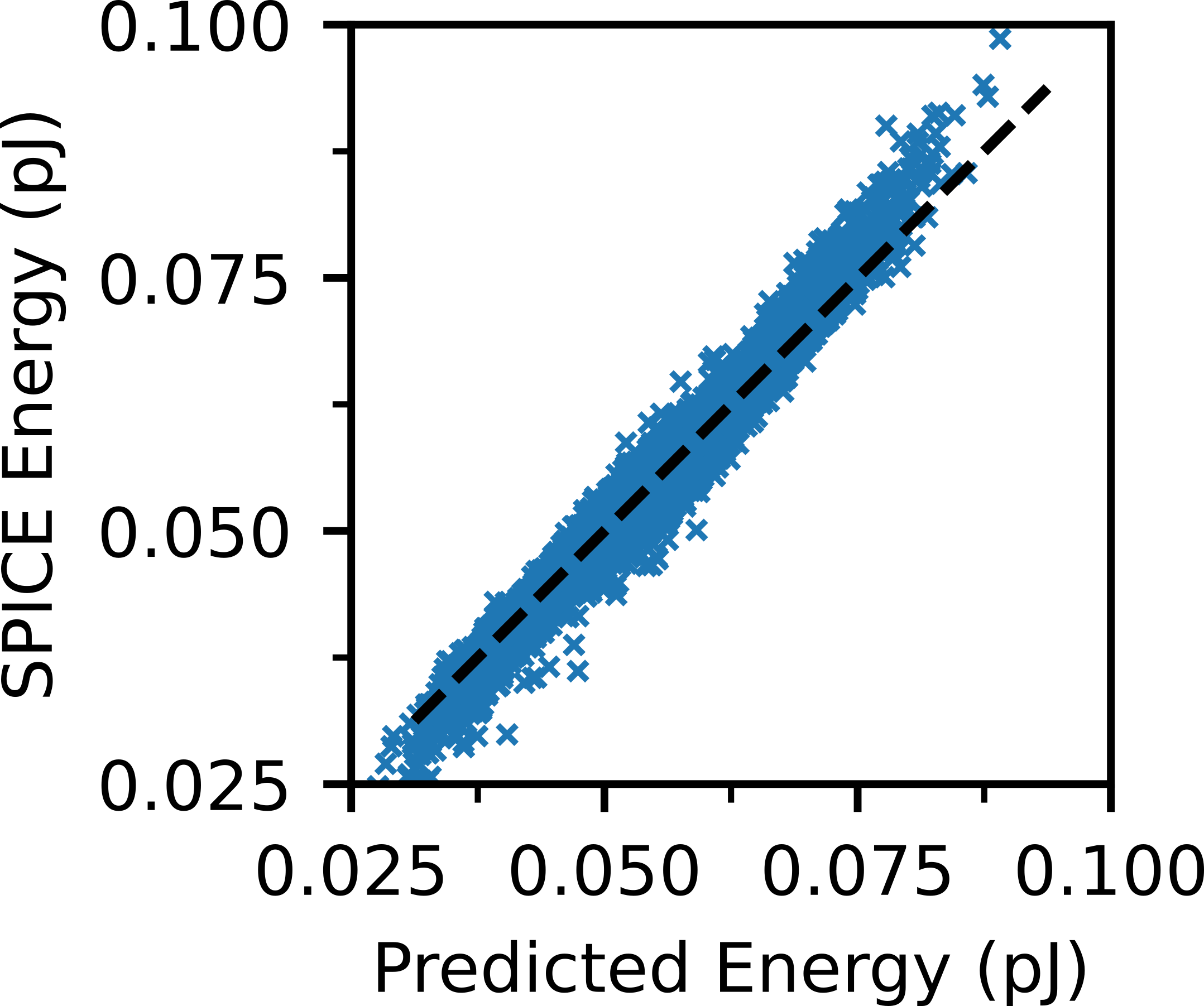}\label{fig:sub11}}
\subfloat[$M_{E_S}$]{\includegraphics[width=0.2225\columnwidth]{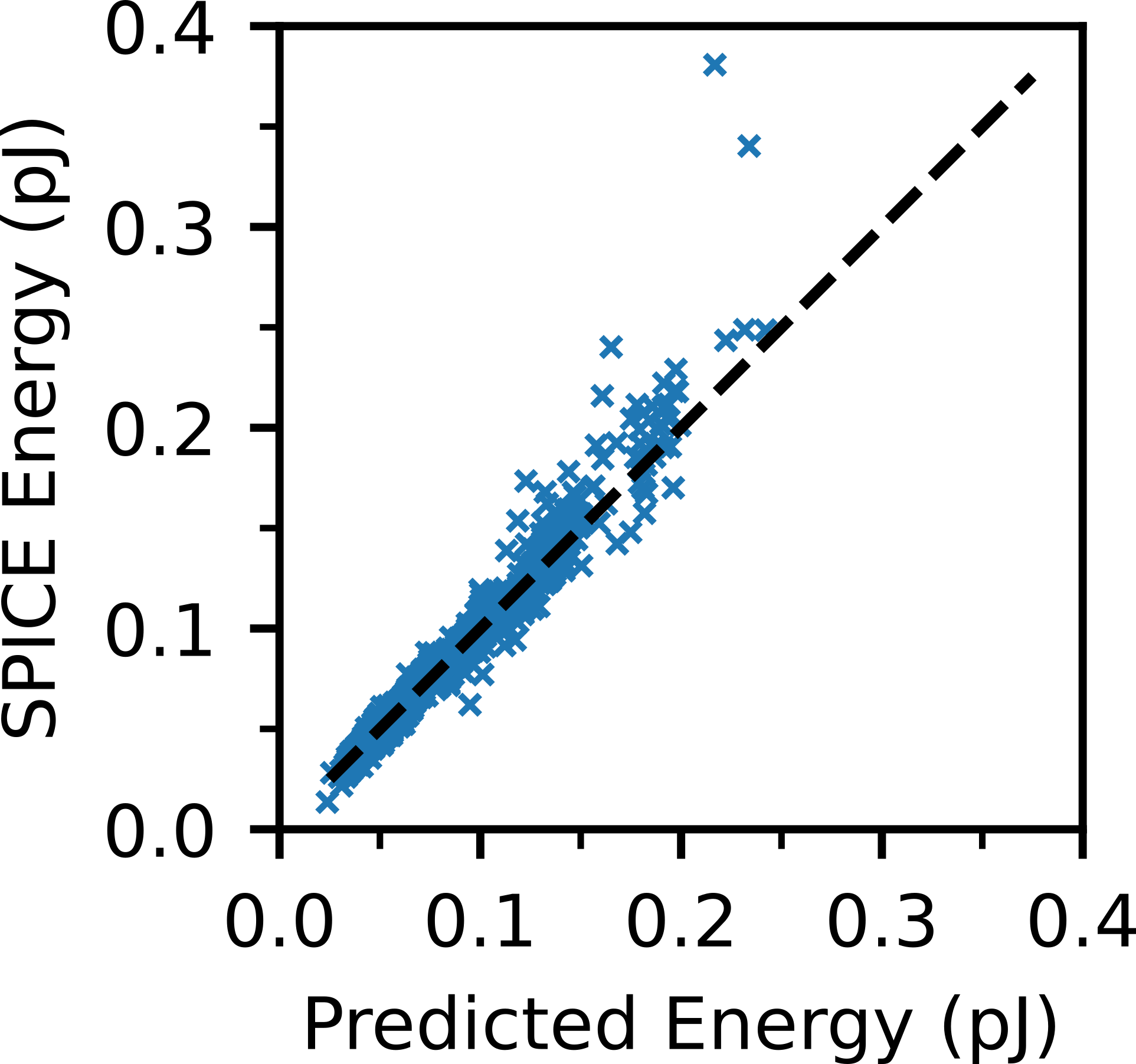}\label{fig:sub22}} 
\subfloat[$M_L$]{\includegraphics[width=0.237\columnwidth]{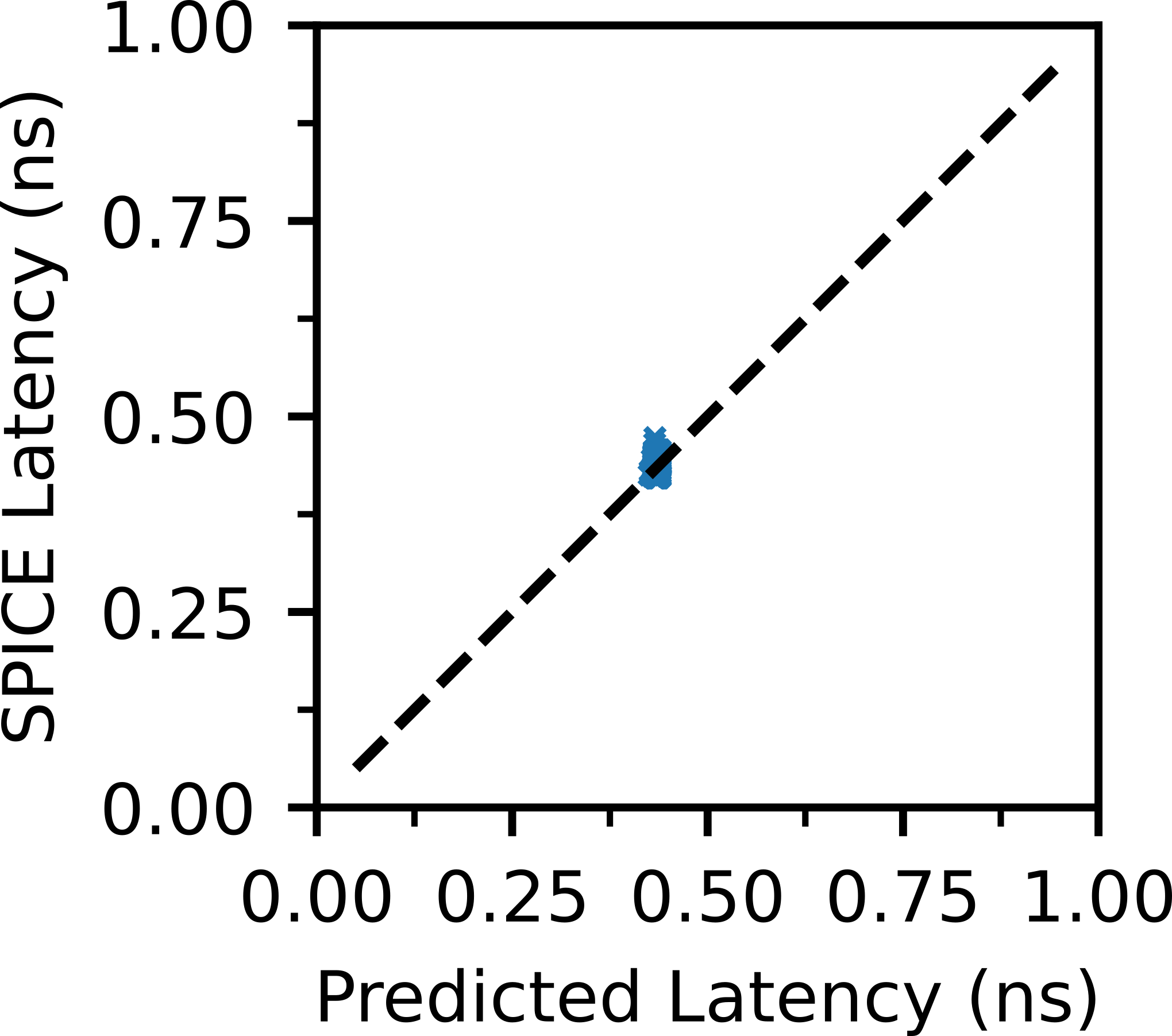}\label{fig:sub33}}
\subfloat[$M_O$]{\includegraphics[width=0.215\columnwidth]{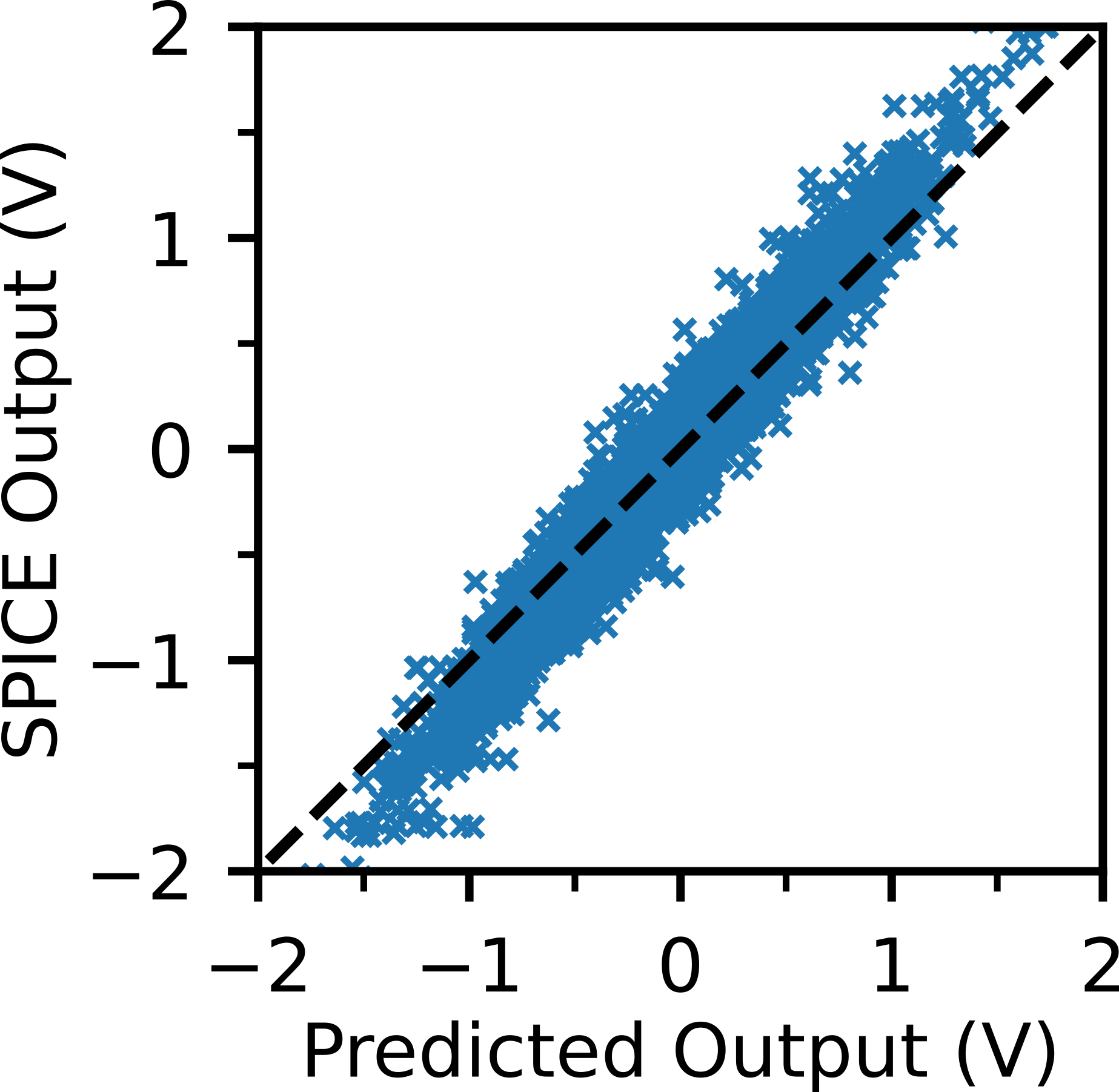}\label{fig:sub44}}
\caption{Correlation plots for the CatBoost model on the crossbar.}
\label{fig:sim-mac}
\end{figure}

To evaluate our models, we use mean squared error (MSE) and mean absolute percentage error (MAPE). We do not report MAPE for value predictors as the metric depends heavily on the underlying data distribution. We also do not report MAPE for static energy since it has small values distributed around zero that overamplify percentage error.

\subsection{Energy and Latency Prediction}
In Table \ref{tab:ml-model-analysis}, we present the test results of the models on the two circuits, with the best performance in each column shown in bold. For the crossbar, the latency distribution is closely clustered around 0.45ns, resulting in similar MSEs and MAPEs across the models. For dynamic energy, CatBoost outperforms other models with a significant margin, achieving 7.3x better MSE and 2.9x better MAPE than the next best MLP. For static energy, the MLP achieves the best MSE, followed closely by CatBoost. Mean, table-based, and linear predictors perform poorly due to high-dimensional inputs. Overall, CatBoost is the best or near-best predictor for the crossbar, which we show in correlation plots in Fig.~\ref{fig:sim-mac}(a)-(c).

In the LIF neuron, energy and latency are functions of both the state and input. For latency, the CatBoost and MLP models achieve similar MSEs, but CatBoost has 1.1x better MAPE. For dynamic energy, we see a similar trend where the MSE of the MLP is 1.4x better than CatBoost, but CatBoost has 1.1x better MAPE. Lastly, for static energy, CatBoost and the MLP perform similarly well. While table-based predictors capture the neuron's complex behavior in a lower-dimensional input space more effectively than linear and mean models, more expressive models like CatBoost and MLP achieve even lower error. Overall, the MLP is the best or near-best predictor for the LIF neuron, which we show in correlation plots in Fig.~\ref{fig:sim}(a)-(c). We note that mispredictions can appear visually exaggerated even when the overall error is low, as in Fig.~\ref{fig:sim}(b).

\subsection{Behavior Prediction}
\begin{figure}[t]
\centering
\vspace{-8mm}
\subfloat[$M_{E_D}$]{\includegraphics[width=0.203\columnwidth]{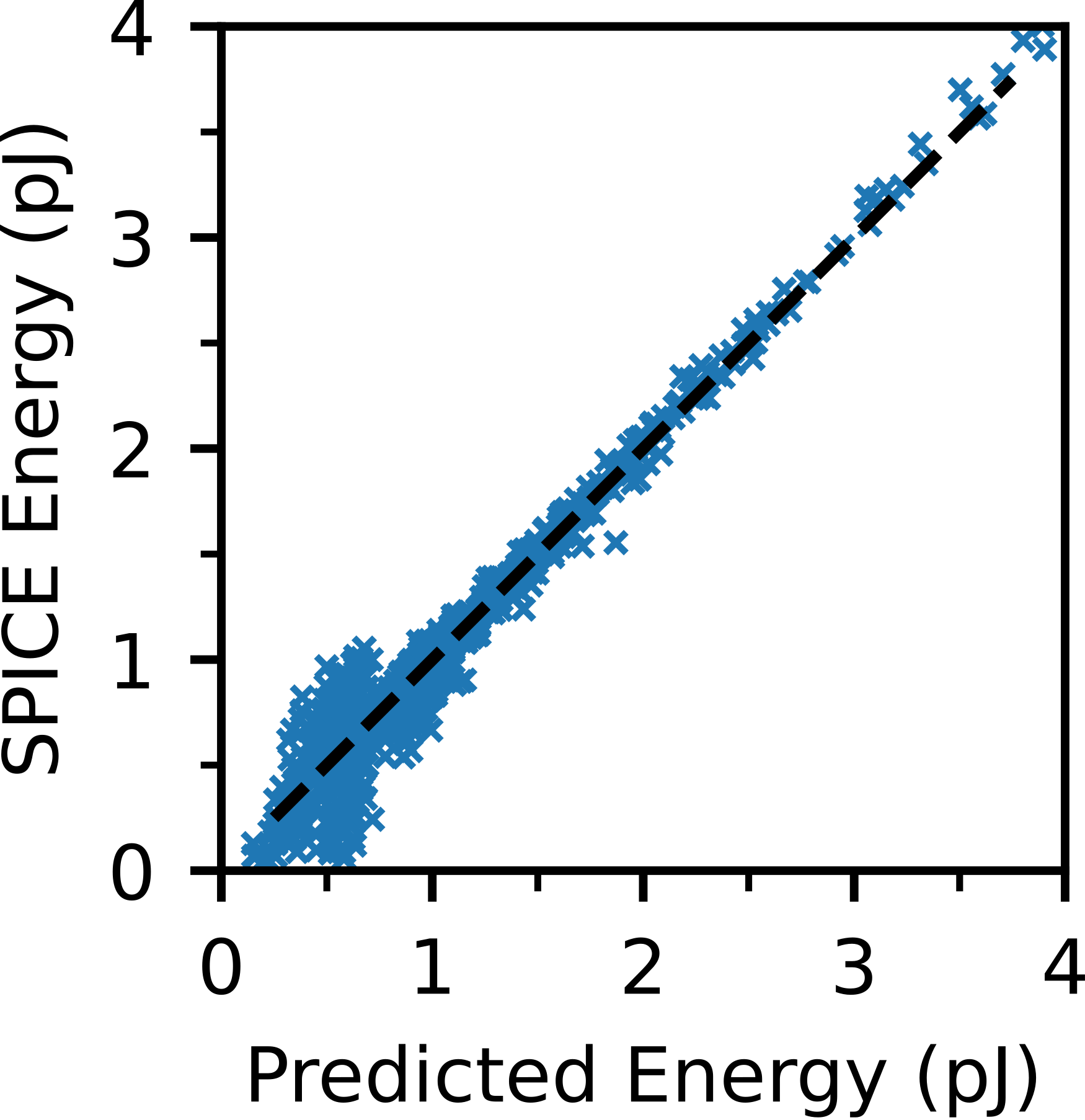}\label{fig:sub1}}
\subfloat[$M_{E_S}$]{\includegraphics[width=0.225\columnwidth]{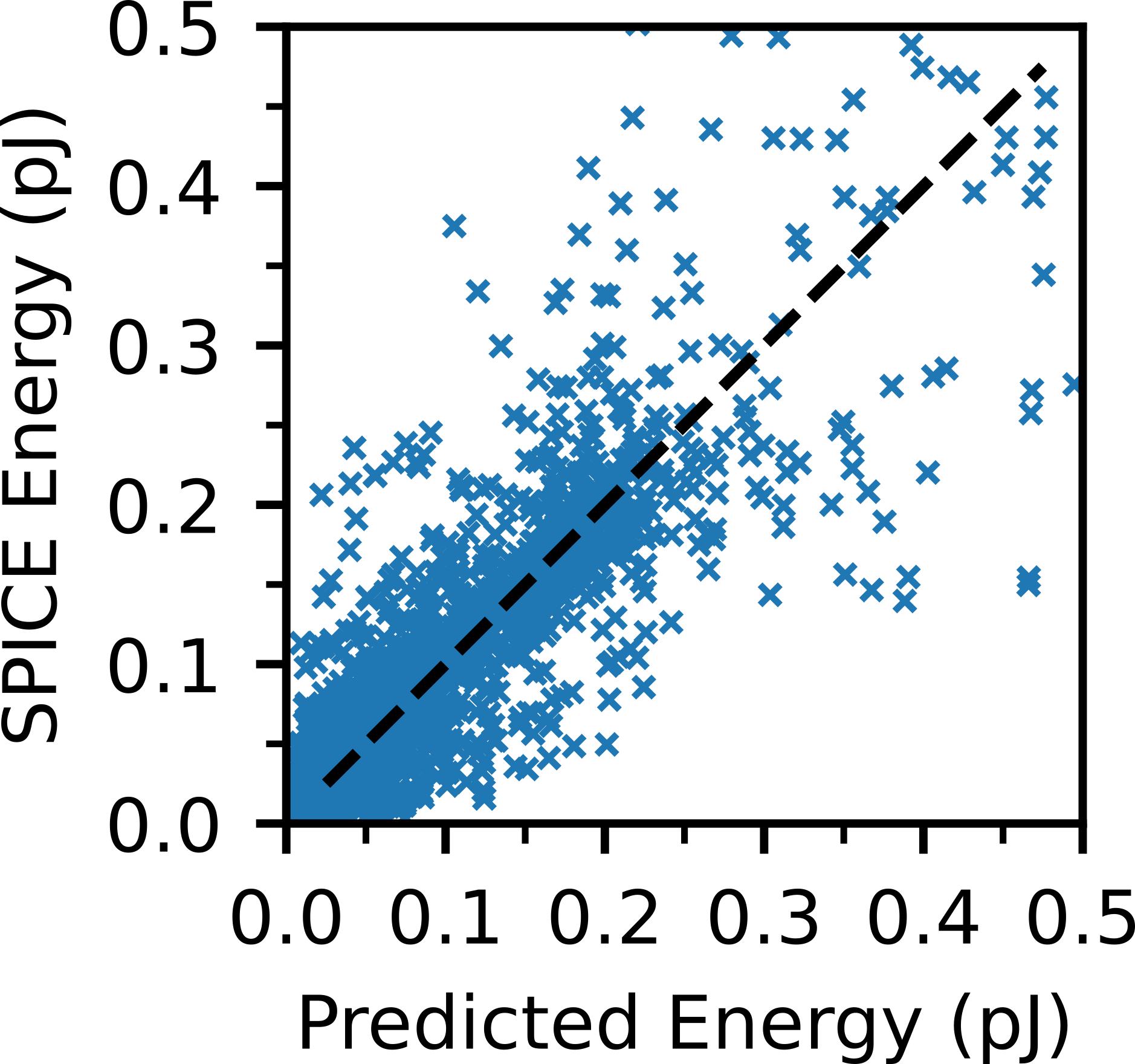}\label{fig:sub2}} 
\subfloat[$M_L$]{\includegraphics[width=0.225\columnwidth]{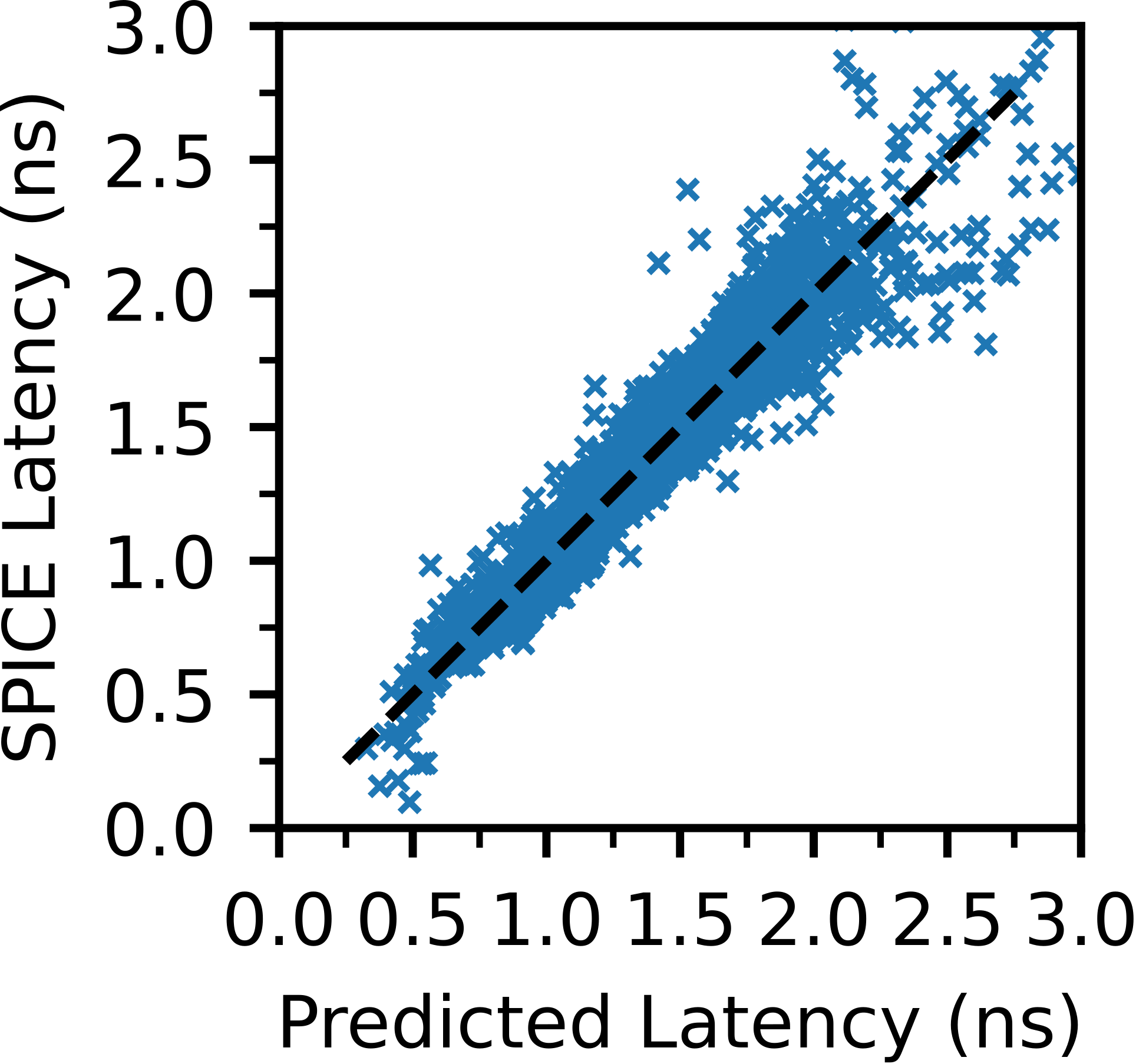}\label{fig:sub3}}
\subfloat[$M_V$]{\includegraphics[width=0.225\columnwidth]{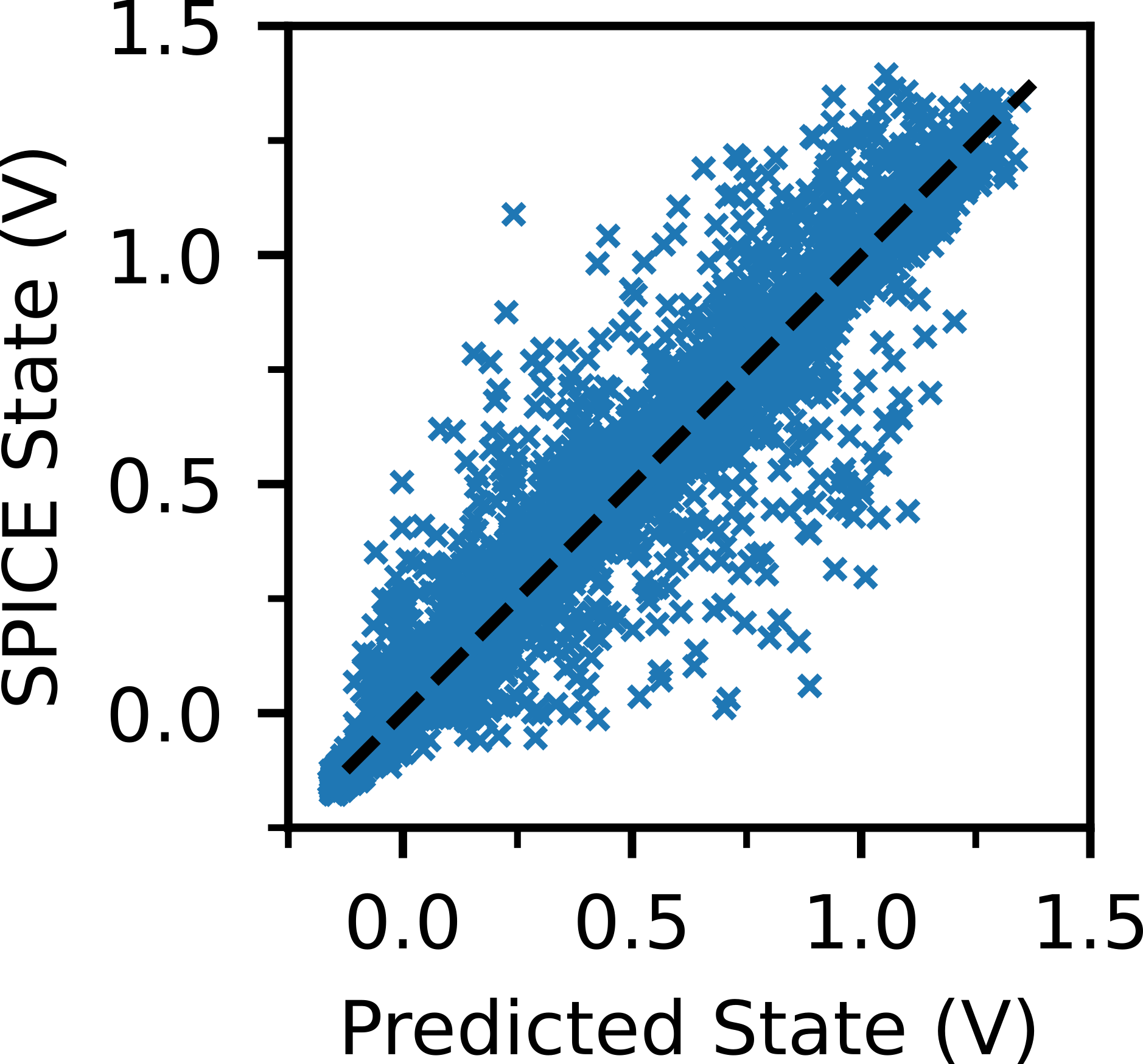}\label{fig:sub4}}
\caption{Correlation plots for the MLP model on the LIF neuron.}
\label{fig:sim}
\end{figure}

Table~\ref{tab:ml-model-analysis} also shows results for state and output prediction. The crossbar output range is $[-2V,2V]$ while the LIF neuron state and output range are $[0V,1.5V]$. For the crossbar, CatBoost achieves the best MSE of 0.0121\,V$^2$ for output prediction. Assuming quantization to 8-bit in the ADC, the MSE reduces to 0.0090\,V$^2$ (1.34x lower). For the LIF neuron, the CatBoost and MLP models perform similarly, with the MLP achieving better MSE for state and output prediction by 1.04x and 1.01x, respectively. When converted to spikes, the MLP's predicted output has 99.3\% accuracy. Again, mean, table-based, and linear predictors perform poorly on the crossbar. With a lower-dimensional input space in the LIF neuron, the table-based model outperforms the mean and linear models, but is still outclassed by more expressive models. We show correlation plots of CatBoost for the crossbar in Fig.~\ref{fig:sim-mac}(d) and MLP for the LIF neuron in Fig.~\ref{fig:sim}(d). Again, mispredictions can appear visually exaggerated as in Fig.~\ref{fig:sim}(d).

\subsection{Behavioral Error Propagation}
In this section, we study the potential for error accumulation as a result of state prediction when using LASANA models. Since the crossbar has no state, predictions have no impact on the future. As such, we focus this analysis on the LIF neuron circuit using the MLP models.

\begin{figure}[t!]
    \centering
    \vspace{-2pt}
    \includegraphics[width=0.48\textwidth]{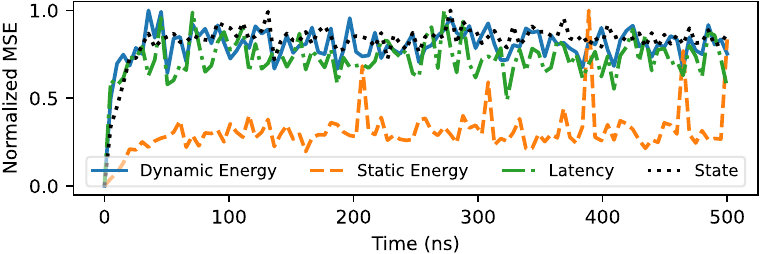}
    \vspace{-8pt}
    \caption{MSE per timestep across a 20,000 neuron layer for 500\,ns.}
    \label{fig:mse-over-time}
    \vspace{-5pt}
\end{figure}

To evaluate the impact of ML-based behavioral modeling, we predict behavior over an entire simulation of a one-layer neural network. Since state predictions in one time step are used as input for subsequent predictions, errors can propagate. Table~\ref{tab:error-prop} shows the results of a study on a 20,000 LIF neuron layer simulated for 500\,ns using randomized circuit parameters and input. LASANA-O is an oracle with perfect knowledge of the state, while LASANA-P uses the predicted state for subsequent prediction. Using LASANA-P, latency MAPE increases by 1.4x, dynamic energy MAPE by 1.3x, and static energy MSE by 2.9x compared to LASANA-O. For state prediction, MSE increased by 3.1x, and output prediction MSE by 1.9x, resulting in an overall decrease in spike accuracy of 0.41\%. When MSE is calculated for all events in a timestep, normalized per predictor, and plotted in Fig.~\ref{fig:mse-over-time}, MSE does not monotonically increase. This shows that ML-based behavioral modeling impacts predictor accuracy, but overall error does not worsen over simulation time.

\begin{table}[t!]
\centering
\caption{Behavioral error propagation for a 20,000 LIF neuron layer.}
\label{tab:error-prop}
\vspace{-5pt}
\setlength{\tabcolsep}{4pt}
\begin{tabular}{@{}lccccccc@{}}
\toprule
                   & \multicolumn{2}{c}{$M_L$}                        & \multicolumn{2}{c}{$M_{E_D}$}                    & $M_{E_S}$                        & $M_V$                           & $M_O$                           \\\cmidrule(lr){2-3} \cmidrule(lr){4-5} \cmidrule(lr){6-6} \cmidrule(lr){7-7} \cmidrule(l){8-8}
                   & \textbf{MSE}                     & \textbf{MAPE} & \textbf{MSE}                     & \textbf{MAPE} & \textbf{MSE}                     & \textbf{MSE}                    & \textbf{MSE}                    \\
\textbf{Model}     & \textbf{(ns\textsuperscript{2})} & \textbf{(\%)} & \textbf{(pJ\textsuperscript{2})} & \textbf{(\%)} & \textbf{(pJ\textsuperscript{2})} & \textbf{(V\textsuperscript{2})} & \textbf{(V\textsuperscript{2})} \\\midrule
\textbf{LASANA-O}    & 0.0160                           & 5.00          & 0.008                            & 7.69          & 0.0021                           & 0.0017                          & 0.0099                          \\
\textbf{LASANA-P} & 0.0364                           & 7.03          & 0.013                            & 9.68          & 0.0060                           & 0.0052                          & 0.0190             \\
\bottomrule
\end{tabular}
\end{table}

\subsection{Runtime Scaling Comparison}
In Table~\ref{tab:sim-times}, we compare the runtimes of Cadence Spectre APS SPICE simulations, purely behavioral SV-RNM models, SV-RNM models with our energy and latency annotation, and standalone ML simulations as the size of an LIF neuron layer increases. Similar runtime trends are observed in the crossbar. Versus SPICE, our speedup factor scales to more than four orders of magnitude at 20,000 neurons. ML-based annotation of energy and latency increased SV-RNM simulation runtime by less than 1.2\% in the 20,000 neuron simulation. By contrast, standalone LASANA was 34.6x faster than SV-RNM. To show LASANA's scaling capability, we also simulated networks of 200,000 neurons and 200,000 32-input crossbar rows, resulting in runtimes of 65.7\,s and 336.1\,s, respectively. For comparison, Intel's Loihi chip has 131,072 neurons \cite{davies_loihi_2018}, and the crossbar rows would be equivalent to 6,250 $32\times32$ crossbar arrays.

\subsection{MNIST and Spiking MNIST Case Study}
Using the crossbar row CatBoost models, we performed a case study on the unpadded $20 \times 20$ MNIST dataset from~\cite{lecun_gradient-based_1998} using a $400 \times 120 \times 84 \times 10$ binary neural network from \cite{amin_imac-sim_2023}. The accelerator partitions the layers into 67 $32\times32$ PCM-based crossbars based on \cite{amin_imac-sim_2023}. To emulate each crossbar, we instantiate 32 instances of the 32-input LASANA crossbar row models. The crossbars perform analog matrix-vector multiplication, where outputs are converted into digital values through an 8-bit ADC, passed through an inverse sigmoid activation layer, and converted back to analog and passed to the next layer through an 8-bit DAC. Running Synopsys HSPICE on the test dataset took 10.96 hours of runtime with 95.86\% MNIST accuracy. By contrast, our models ran in 0.13 hours (82x speedup) with MNIST accuracy of 97.61\% (1.75\% error), average per-inference dynamic energy and latency MAPE of 2.44\% and 2.65\%, and average total energy and latency error per inference of 0.8\% and 2.55\%, respectively.

\begin{table}[t!]
\centering
\caption{500\,ns simulation runtimes with varying LIF neuron layer size.}
\vspace{-5pt}
\label{tab:sim-times}
\setlength{\tabcolsep}{5pt} 
\begin{tabular}{@{}cccccc@{}}
\toprule
\multicolumn{1}{l}{} & \multicolumn{3}{c}{\textbf{Runtime (s)}}                               & \multicolumn{2}{c}{\textbf{Speedup}} \\\cmidrule(lr){2-4} \cmidrule(l){5-6}
\textbf{Neurons}     & \textbf{SPICE} & \textbf{SV-RNM (+ ML)} & \textbf{Ours} & \textbf{SPICE}   & \textbf{SV-RNM}   \\ \midrule
10                   & 3.50               & 0.41 + 0.01                 & 0.04              & 85.4             & 10.0              \\
100                  & 50.29              & 0.67 + 0.05                 & 0.08              & 613.5            & 8.2               \\
1000                 & 1246.15            & 3.23 + 0.08                 & 0.18              & 6736.6           & 17.5              \\
3000                 & 4101.40            & 9.13 + 0.16                 & 0.38              & 10735.6          & 23.9              \\
5000                 & 7464.89            & 14.98 + 0.23                & 0.57              & 13104.6          & 26.3              \\
20000                & 31681.10           & 63.39 + 0.75                & 1.83              & 17275.2          & 34.6             \\
\bottomrule
\end{tabular}
    \vspace{-5pt}
\end{table}

Using the LIF neuron MLP models, we perform a similar study on the $28\times28$ padded MNIST dataset from~\cite{lecun_gradient-based_1998}, where we convert each pixel into a Poisson rate-encoded spike train with firing probability proportional to pixel intensity. A $784\times128\times10$ SNN was trained using SNNTorch~\cite{eshraghian_training_2023} with the Adam optimizer and MSE count loss that encourages 60\% spike activity on the correct output neuron and 20\% on the incorrect neurons. Each inference ran for 100 timesteps (500\,ns). All LIF neuron tunable parameters were set to 0.5\,V except $V_{leak}$ at 0.58\,V. To emulate the SNN, we instantiate a LASANA instance for each neuron with connectivity based on the network. Running Cadence Spectre APS on spiking MNIST took 2404 hours with 95.65\% MNIST accuracy. By contrast, MLP LASANA models ran in 1.82 hours (1321x speedup) with MNIST accuracy of 95.63\% (0.02\% error), average per-inference dynamic energy and latency MAPE of 6.92\% and 7.58\%, and average total energy and latency error per inference of 2.90\% and 5.88\%, respectively.

Results show that LASANA produces fast and accurate models for full-system simulations of a wide variety of hybrid analog/digital architectures. In all cases, application accuracy slightly decreases from baseline results due to prediction errors propagating through successive network layers. The spiking model achieved less behavioral error due to the binary nature of spikes, which masks the propagation of small variations.

\section{Summary and Conclusions}
In this paper, we propose LASANA, an automated ML-based approach for surrogate modeling to provide energy and latency annotation of behavioral models or completely replace analog circuit simulations for large-scale neuromorphic architecture exploration. Our models are general to support a wide range of circuits. To show the flexibility of our methodology, we apply it to a memristive crossbar array and a spiking LIF neuron. Experimental results for MNIST demonstrate up to three orders of magnitude speedup over SPICE, with energy, latency, and behavioral error less than 7\%, 8\%, and 2\%, respectively. In future work, we aim to integrate LASANA into existing digital simulators~\cite{boyle_performance_2023,SANA-FE}, apply it to a wider range of circuits, support device variability, and explore LASANA models for circuit-aware application training.

\pagebreak

\bibliographystyle{IEEEtran}
\bibliography{references.bib}
\end{document}